# Study of the crystal structure, superconducting and magnetic properties of Ru$_{1-x}$Fe$_x$Sr$_2$GdCu$_2$O$_8$

R. Escamilla, F. Morales, T. Akachi, and R. Gómez[*]

Instituto de Investigaciones en Materiales, Universidad Nacional Autónoma de México, C. U., Apdo. Postal 70-360, México, D. F., 04510, México.
*Facultad de Ciencias, Universidad Nacional Autónoma de México, C. U. Circuito Exterior, México, D. F., 04510, México.

**Abstract**

Samples of the Ru$_{1-x}$Fe$_x$Sr$_2$GdCu$_2$O$_8$ system with x = 0, 0.025, 0.05, 0.075, 0.1, and 0.2, were prepared and their structural, superconducting and magnetic properties were studied. Rietveld refinement of the X-ray diffraction patterns show that the Fe substitution occurs in both Ru and Cu sites. An increase of Fe concentration produces no significant changes in the bond angle Ru-O(3)-Ru, which is a measure of the rotation of the RuO$_6$ octahedra around the c-axis, and also in the bond angle ϕ, (Ru-O(1)-Cu), which is a measure of the canting of the RuO$_6$ octahedra. On the other hand, the bond angle Cu-O(2)-Cu, which is a measure of the buckling of the CuO$_2$ layer, has a slight tendency to decrease with the increase of Fe content. We found that both ferromagnetic and superconducting transition temperatures are reduced with the increase of Fe concentration. Analysis related to the decay of the superconducting and ferromagnetic states is presented.

**Keywords:** Ru1212; X-ray diffraction; Crystal structure; ferromagnetism, Rietveld, Substitution effect

**PACS:** 74.62.Bf, 74.62.Dh, 74.72.Bk

## 1. Introduction

The discovery of the coexistence of superconductivity and ferromagnetism in the ruthenocuprate compound $RuSr_2GdCu_2O_8$ (Ru-1212) [1-5] has raised considerable interest in understanding the intrinsic properties of this layered material. The tetragonal crystal structure of the Ru-1212 compound can be described based on the similarity to $REBa_2Cu_3O_{7-\delta}$ (RE-123) superconductors. The structure of Ru-1212 contains two $CuO_2$ layers separated by a single oxygen-less Gd layer, the $RuO_2$ layer replacing the CuO chains present in RE-123 superconductors and a SrO layer located between the $CuO_2$ and $RuO_2$ layers. The superconductivity is associated with the $CuO_2$ layers, as in the RE-123 superconductors, while the ferromagnetism seems to be induced in the $RuO_2$ layers. The ferromagnetic transition temperature, $T_M$, is about 135 K, while the superconducting transition temperature, $T_c$, occurs in the 0-45 K range, depending on sample preparation procedures [6,7]. Zero-field muon spin rotation measurements [3] and other experiments have shown [8,9] that the Ru-1212 compound is microscopically uniform with no evidence for spatial phase separation of superconducting and magnetic regions, indicating the three dimensional character of superconductivity and a uniform long-range magnetic order. Neutron diffraction experiments [10,11] have demonstrated that the Ru sublattice shows a G-type antiferromagnetic structure, with an ordering moment of the order of 1 $\mu_B$. From magnetization measurements a ferromagnetic ordering has been observed and it was proposed that the origin of the ferromagnetic moment is the canting of Ru moments that give a net moment perpendicular to the c-axis [11,12]. However, how a ferromagnetic component emerges from an antiferromagnetic background is still unclear.

Several studies of cation substitutions in the Ru-1212 compound have been reported in the literature [13-20]. Their effects in the superconducting and magnetic properties depend on the type of cation and the substitutional site. Studies in the $Ru_{1-x}Sn_x$-1212 system [13,14] show that the Sn doping suppresses the ferromagnetic moment in the $RuO_2$ layer, decreasing $T_M$ but, on the other hand, $T_c$ increases with the increase of Sn content. These results were attributed to the diamagnetic properties of Sn ions that reduce the total magnetic moment in the $RuO_2$ layers and increase the hole transfer to the $CuO_2$ layers. Substitutions in Ru sites by Ti [15,16], Nb [15], Rh [16], and Co [17] have shown that both $T_M$ and $T_c$ decrease with increasing doping. A peculiar behavior of enhancing both $T_M$ and $T_c$ has been found by V substitution in Ru sites [15]. This behavior could be attributed to the ability of vanadium to adopt a 4+ / 5+ mixed-valence. Klamut et al. [18] have investigated Cu substitutions in Ru sites. They found that $T_c$ values strongly increase with doping reaching a maximum of 72 K and when Ru is substituted by Cu up to 20 at. %, they detected signals of magnetic ordering, above the superconducting transition temperatures, but no ferromagnetic signals of the Ru sublattice were detected at higher doping levels. They also observed a reentrant magnetization below $T_c$ due to the paramagnetic response of the Gd sublattice. Studies of La substitution in Sr sites [19] have been performed and the results show that $T_M$ increases slightly, but superconductivity is strongly reduced by doping, due to a hole trapping mechanism by disorder defects. Substitution of Zn in Cu sites [20], as in others high-$T_c$ cuprate superconductors, rapidly suppress the superconductivity due to pair breaking mechanisms.

Although there is general agreement that in the $RuSr_2GdCu_2O_8$ the superconductivity is originated in the $CuO_2$ layers and the ferromagnetism in the $RuO_2$ layers, further understanding is required to know about the nature of this ferromagnetic superconductor. With this objective in mind, we performed Fe doping experiments, asking about the possible influence of the iron magnetic moment on the magnetic and superconducting properties of the Ru-1212 compound. Thus, in this paper we report our results related to the structural properties, the electrical resistivity and the magnetic properties of the $Ru_{1-x}Fe_xSr_2GdCu_2O_8$ system as a function of Fe doping, temperature and external magnetic field.

**2. Experimental details**

Polycrystalline samples of $Ru_{1-x}Fe_xSr_2GdCu_2O_8$ ($Ru_{1-x}Fe_x$-1212), with x = 0, 0.025, 0.05, 0.075, 0.1 and 0.2, were synthesized by solid state reaction of stoichiometric quantities of oxides of $RuO_2$ (99 %), $Fe_2O_3$ (99.999 %), $Gd_2O_3$ (99.9 %), CuO (99.99 %), and $SrCO_3$ (98+ %). After calcination in air at 900 °C, the material was ground, pressed into pellets and annealed in oxygen atmosphere at 1000 °C for 72 hours. Phase identification of the samples was done with a X-ray diffractometer Siemens D5000 using Cu-$K_\alpha$ radiation and a Ni filter. Intensities were measured at room temperature in steps of 0.02°, for 14 seconds, in the 2θ range 20° - 100°. The crystallographic phases were identified by comparison with the X-ray patterns of the JCPDS database. The crystallographic parameters were refined using a Rietveld refinement program, Rietica v 1.7.7 [21] with multi-phase capability. The superconducting transition temperatures were determined in a closed-cycle helium refrigerator by measuring resistance *vs*. temperature. The resistance was measured by the four-probe technique in the temperature range of 14 K to 250 K. dc-magnetization measurements were performed in a superconducting quantum interference devise (SQUID) based magnetometer, in the temperature range of 2 K to 300 K.

3. **Results and discussion**

Figure 1 shows the X-ray diffraction patterns for the synthesized samples of $Ru_{1-x}Fe_x$-1212. The analysis of these data indicate that the crystal structure of the samples correspond to that of Ru-1212 structure, although for x = 0 faint features of the $SrRuO_3$ structure (ICDD n° 41-1442) were observed and additional peaks corresponding to the $Sr_3(Ru,Cu)_2O_7$ (ICDD n° 51-0307) phase were also detected for x ≥ 0.025. The X-ray diffraction patterns of the samples were Rietveld-fitted using a space group P4/mmm (n° 123), taken into account the possibility that Fe can also occupy Cu sites and the presence of $SrRuO_3$ and $Sr_3(Ru,Cu)_2O_7$ secondary phases. As an example, we show in Fig. 2 the fitted pattern of the X-ray spectra for the undoped sample.

The structural parameters obtained from the Rietveld refinements are shown in Table 1. The oxygen atoms localized in the SrO layer are denoted as O(1), those located in the $CuO_2$ layer as O(2) and by O(3) those in the $RuO_2$ layer. N(Fe) represents the occupancy parameter for Fe in the Ru and Cu sites. From the refinement results it is clear that the Fe ions occupy both the Ru and Cu sites. The Table shows the crystallographic parameter values for all the samples studied; the values determined for the undoped sample are in agreement with other published results [5, 10]. Figure 3 shows the lattice parameters and the cell volume of the samples as a function of iron content x. The a-axis shows a slight increase with increasing x, while the c-axis shows a significant decrease with x. The net result is a decreasing volume with increasing x. A list of the Rietveld-fitted bond lengths and bond angles of the samples is shown in Table 2. We observe that both the Ru(Fe) – O(1) and Cu(Fe) – O(1) bond lengths decrease with increasing Fe content and they should be associated with the shortening of the c axis.

The different characteristic angles of the structure show the following behavior with increasing x: (a) The bond angle $\phi$ (Ru-O(1)-Cu), which is related to the deviation of the apical oxygen O(1) along the plane perpendicular to the c-axis, whose value determines the distortion of the $RuO_6$ octahedra, essential for the magnetic exchange interaction, shows no significant changes. (b)The bond angle Ru-O(3)-Ru, which is a measure of the rotation of the $RuO_6$ octahedra around the c-axis, remains constant. (c) The bond angle Cu-O(2)-Cu, which is a measure of the buckling of the $CuO_2$ layer, shows a slight tendency to decrease indicating an increase in the buckling of the $CuO_2$ layer.

Figure 4 shows the normalized resistance as a function of temperature for all investigated samples. The R(T) curves for samples showing superconducting transitions are plotted in the inset of the figure. On decreasing temperature, the resistance curve of the undoped sample shows a steady decrease until a relative minimum is attained around T = 75 K; then a slight increase is observed just before the onset of superconductivity at T = 45 K; zero resistance state is reached at T = 25 K. For the x = 0.025 sample, the R(T) curve increases as temperature decreases until it reaches the onset of superconducting transition temperature at T = 30 K. For samples with x ≥ 0.05, the R(T) curves show a semiconducting-like temperature behavior, without any signal of a superconducting transition, at least to the minimum temperature of 14 K investigated.

The superconducting transition temperature of the $Ru_{1-x}Fe_x$-1212 system drops quite fast with the increase of iron content and superconductivity is suppressed around 5 % of Fe substitution. From the Rietveld refinement results, the Fe atoms partially substitute Cu atoms in the $CuO_2$ layers, and many previous studies in high-$T_c$ cuprate superconductors have shown that Fe substitution in $CuO_2$ layers rapidly degrade the superconducting state [22-25].

Furthermore, some studies have shown a correlation between the buckling of the $CuO_2$ layers and the superconducting transition temperature [26]. The highest $T_c$ is achieved in structures with flat and square $CuO_2$ layers and long apical Cu-O bond lengths. In other words, an increase in the buckling of the $CuO_2$ layers and the shortening of the apical Cu-O bond lowers the transition temperature, due to a related hole localization phenomenon. Our Rietveld refinement studies show that there is an increasing buckling tendency of the $CuO_2$ layer and a decreasing Cu-O(1) bond length with the increase of Fe content, resulting in a fast degradation of $T_c$ and the ultimate disappearance of superconductivity.

Figure 5 shows the temperature dependence of the field cooled (FC) dc-magnetization measurements, M(T), performed in an applied magnetic field $H_a$ = 100 Oe. All samples show the characteristic ferromagnetic ordering transition curves and, in particular, the transition temperature for the undoped sample occurs at around T = 140 K. As Fe content is increased, a broadening of the transition and a reduction of the magnetic ordering transition temperature is observed. An important result that can be draw out from the Rietveld refinement of our X-ray data is that there are no significant changes in the bond angle $\phi$, which is a measure of the canting of the Ru magnetic moment necessary to explain the appearance of ferromagnetism. Therefore, the above mentioned magnetic behavior is not due to a structural change, but to a weakening of the magnetic interaction between the Ru magnetic moments of the Ru sublattice. This behavior was also observed when Ru is substituted by other elements [13-17]. At temperatures below 40 K the M(T) curves show an increase due to the paramagnetic contribution of the Gd ions. The Gd sublattice remains in the paramagnetic state when the Ru sublattice is ordered ferromagnetically, and orders antiferromagnetically at $T_N$ = 2.8 K [19]. However it is worth to note that in a recent study, the possibility that the net ferromagnetic moment of the Ru sublattice can induce a small component of ferromagnetic ordering at the Gd sites that would contribute to the total magnetic moment of the system[27].

To determine the effect of Fe doping on the effective magnetic momenta, $\mu_{eff}$, as well as on the magnetic ordering temperature of our samples, we fitted the magnetic susceptibility data, $\chi(T) = M(T)/H_a$, with the sum of two Curie-Weiss functions $\chi = C/(T - T_{CW})$, assuming independent contributions of the Gd and the Ru sublattices. The fitting was done in the 150 K to 300 K temperature range. The magnetic parameters of the Gd ions were kept fixed at $\mu_{eff}$ = 7.94 $\mu_B$ and $T_{CW}$ = − 4 K values [19]. The effective magnetic moment was determined trough the relation $C = N\mu^2_{eff}/3k_B$ (N: Abogadro´s number, $k_B$: Boltzmann´s constant). As a representative fit, the inset of Fig. 4 shows, in a continuous line, the $\chi(T)$ of the sample for x = 0.075. The resulting fitting parameters for the Ru sublattice are giving in Table 3 and Figure 6 shows the $\mu_{eff}$ and $T_{CW}$ values as functions of Fe content. We observe that both $\mu_{eff}$ and $T_{CW}$ decrease with the increase of Fe content, results that were expected from our analysis of the weakening of the magnetic interaction between the Ru moments provoked by Fe doping. The value of $\mu_{eff} = 2.87 \mu_B$, obtained for x = 0, is in agreement with other reported values [16].

Magnetization measurements as a function of the applied magnetic field, below the magnetic transition temperatures, were also done. Before each measurement the sample was warmed above 150 K and cooled down in zero field. As an example, Figure 7 shows the M(H) hysteresis curve for the x = 0.05 sample, measured at T = 5 K, with the magnetic field changing between −10 kOe and +10 kOe. The magnetization curve shows a hysteresis loop of ferromagnetic materials although it displays a non saturated characteristic due to the paramagnetic contribution of the Gd sublattice.

Finally we would like to point out that Mössbauer studies, currently in process, have shown no magnetic signals due to the presence of impurities phases detected in the X-ray spectra, so it must be concluded that the obtained magnetic behavior results are only due to the Ru and Gd sublattices of the $Ru_{1-x}Fe_x$-1212 system. Detailed study will be published in the near future.

## 4. Conclusions

We have presented a study of the structural, superconducting and magnetic properties of the $Ru_{1-x}Fe_xSr_2GdCu_2O_8$ system. We found that the superconducting transition temperatures $T_c$ and the ferromagnetic transition temperatures $T_M$, both, decrease with the increase of Fe content. The fast $T_c$ decrement and the consequent quite fast disappearance of the superconducting state were explained by the fact that Fe ions not only occupy Ru sites but also Cu sites. It is well known that ion substitutions in the $CuO_2$ layers of high-$T_c$ superconducting cuprates are a big source of $T_c$ degradation. From the obtained structural changes, the shortening of the apical Cu-O bond length and a tendency of the $CuO_2$ layer to increase buckling with increasing Fe content, could be associated with the observed $T_c$ decrement. No significant changes were observed, with Fe doping, in both the canting bond angle $\phi$ and in the $RuO_6$ octahedra rotation bond angle Ru-O(3)-Ru. Therefore the observed broadening of the transition and the reduction of the magnetic ordering temperature as Fe content is increased are not associated with structural changes, but with a weakening of the magnetic interaction between the Ru moments of the Ru sublattice by doping.

Fitting two Curie-Weiss functions to the magnetic susceptibility data, assuming independent contributions of the Gd and the Ru sublattices, we quantify the effective magnetic momenta, $\mu_{eff}$, of the Ru sublattice and the magnetic ordering temperature, $T_{CW}$, as a function of Fe content.

**Acknowledgments**

The authors want to acknowledge A. Arevalo for his assistance in the preparation of the samples used in this work.

**Figure Captions**

Fig. 1 X-ray diffraction patterns for the $Ru_{1-x}Fe_xSr_2GdCu_2O_8$ samples. The symbols + and * point out peaks of $SrRuO_3$, and $Sr_3(Ru,Cu)O_7$ impurity phases, respectively.

Fig. 2 Rietveld refinement of the X-ray diffraction pattern for the x = 0.0 sample. Experimental spectrum (dots), calculated pattern (continuous line), their difference (middle line) and the calculated peak positions (bottom).

Fig. 3. Crystal lattice parameters and unit cell volume as a function of Fe content x.

Fig. 4 Normalized resistance as a function of temperature of the $Ru_{1-x}Fe_x$-1212 samples. The inset shows the curves for x=0.0 and 0.025 samples, where the superconducting transition is more clearly distinguish.

Fig 5 FC magnetization measurements as function of temperature for the $Ru_{1-x}Fe_xSr_2GdCu_2O_8$ samples. The inset shows the Curie-Weiss fitting of the susceptibility data for the x=0.075 sample. The indicated $\mu_{eff}$ and $T_{CW}$ values correspond to the magnetic parameters of the Ru sublattice.

Fig 6 (a) Curie-Weiss temperature as a function of the Fe content x. Fig 6 (b) Effective magnetic moment as a function of the Fe content x.

Fig. 7 Hysteresis loop for $Ru_{0.95}Fe_{0.05}Sr_2GdCu_2O_8$ sample, measured at 5 K

**Table Captions**

Table 1 Structural parameters for $Ru_{1-x}Fe_x$-1212 at 295 K. Space group: P4/mmm (n. 123). Atomic positions: Ru: 1b (0, 0, 1/2); Gd: 1c (1/2, 1/2, 0); Sr : 2h (1/2, 1/2, z); Cu: 2g (0, 0, z); 2O(1) in 8s (x,0,z) × 1/4, 4 O(2) in 4i (0,1/2,z), and 2O(3) in 4o (x,1/2,1/2) × ½ position. N(Fe) is the iron occupancy parameter.

Table 2 Bond lengths (Å) and bond angles (deg) for $Ru_{1-x}Fe_x$-1212.

Table 3 Parameters obtained from the temperature dependence of the magnetic susceptibility of $Ru_{1-x}Fe_x$-1212. C is the Curie-Weiss constant, $T_{CW}$ is the Curie-Weiss temperature, and $\mu_{eff}$ is the effective magnetic moment associated to Ru.

Figure 1

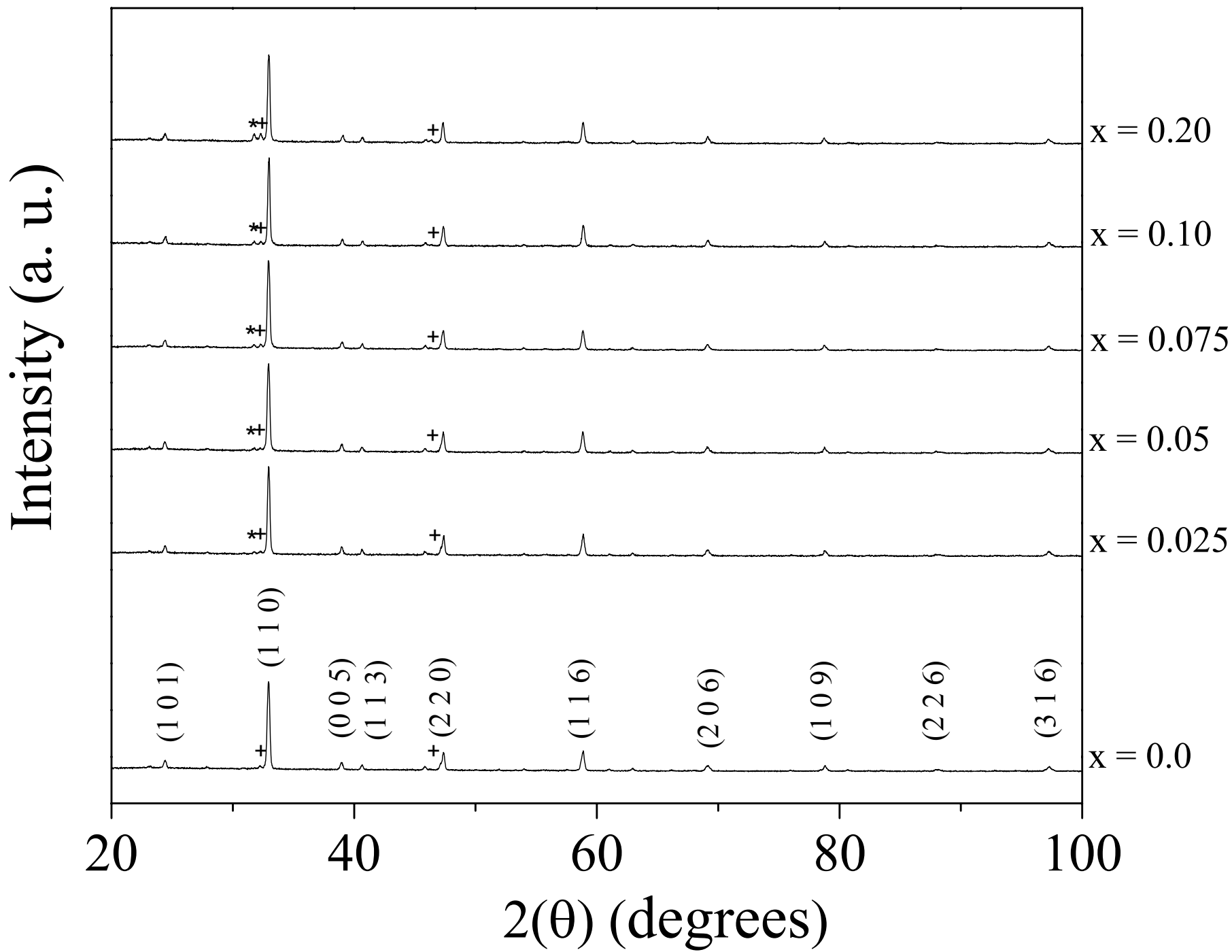

Figure2

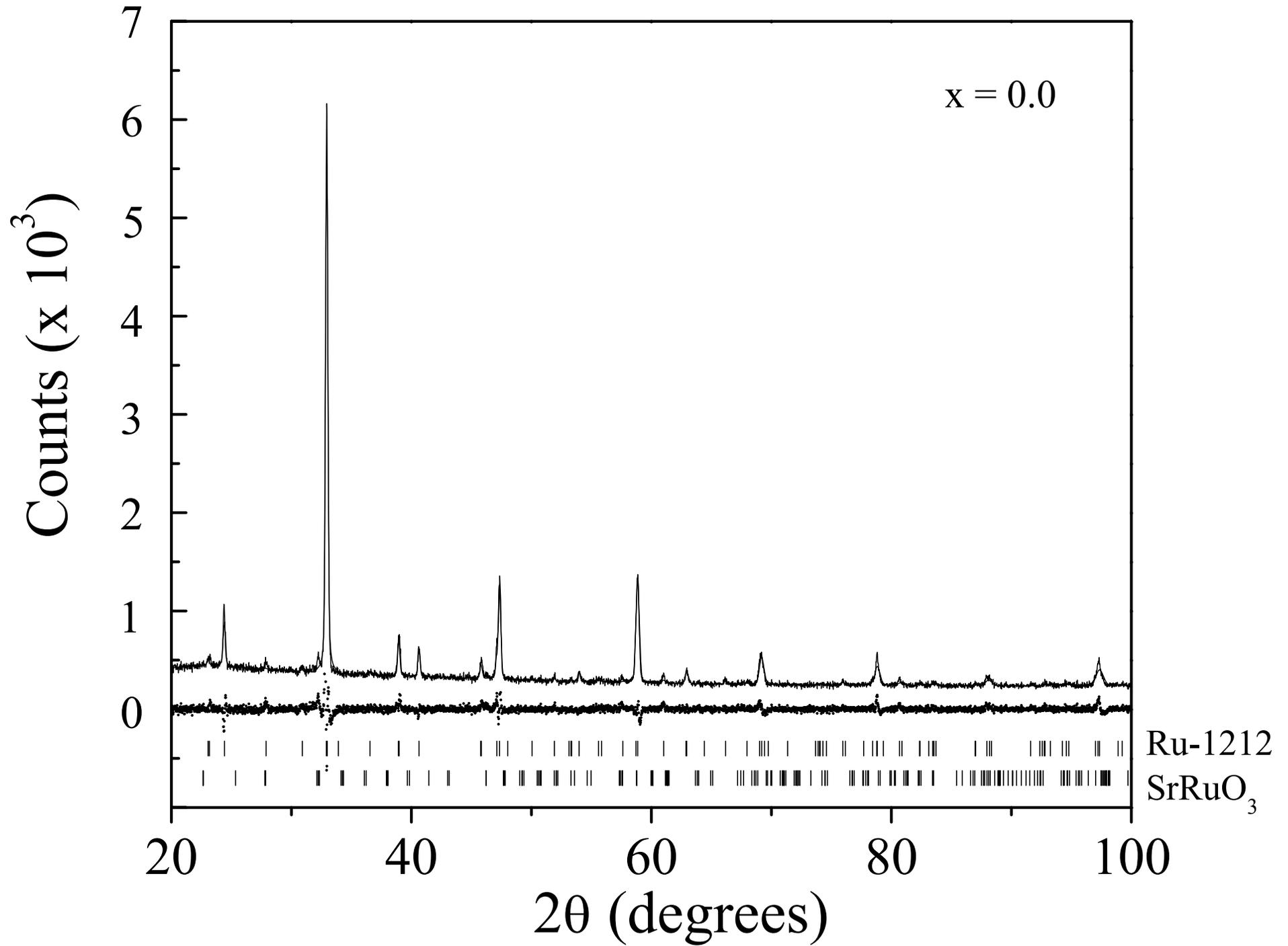

Figure 3

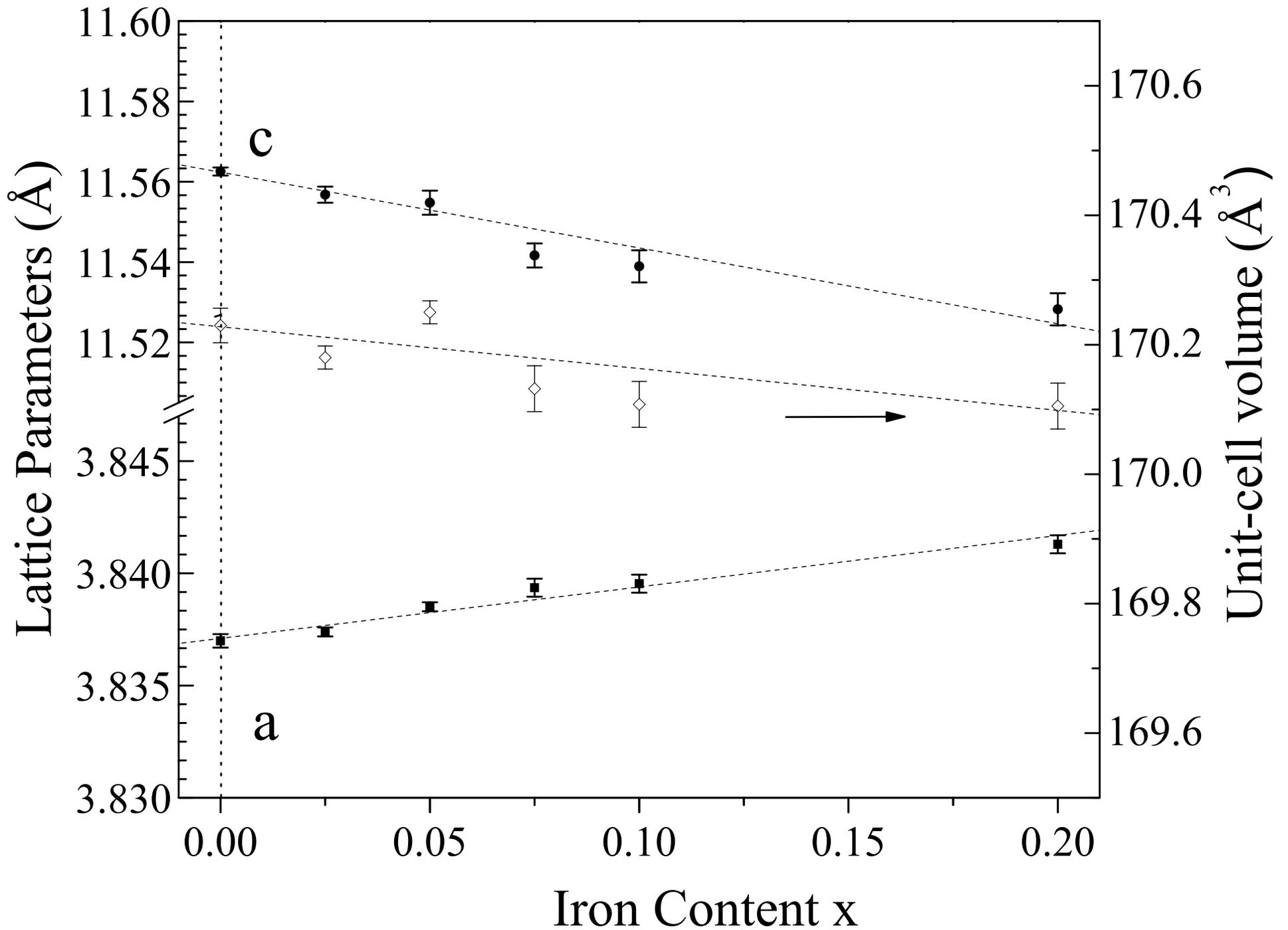

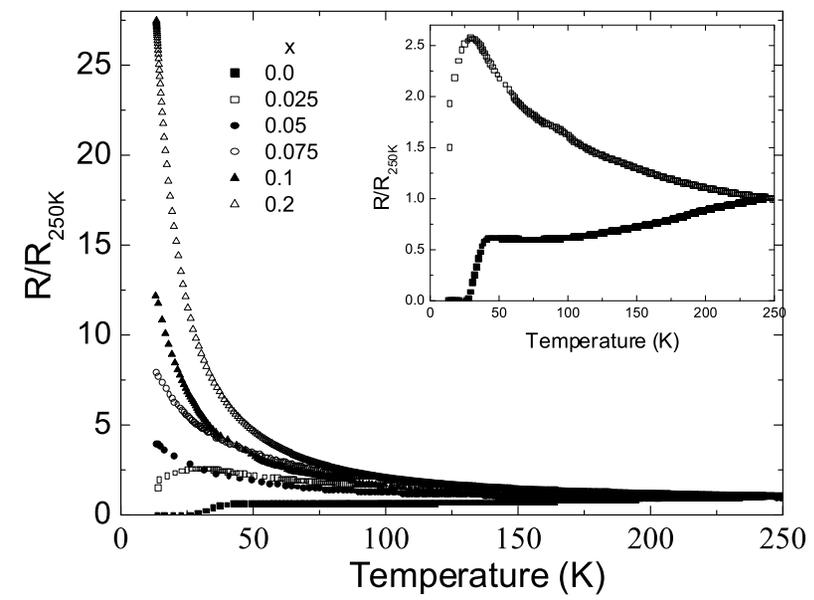

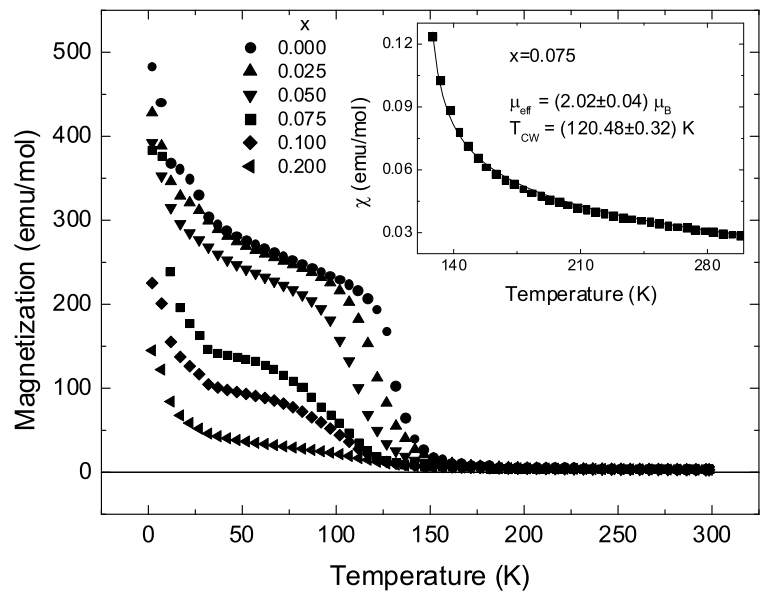

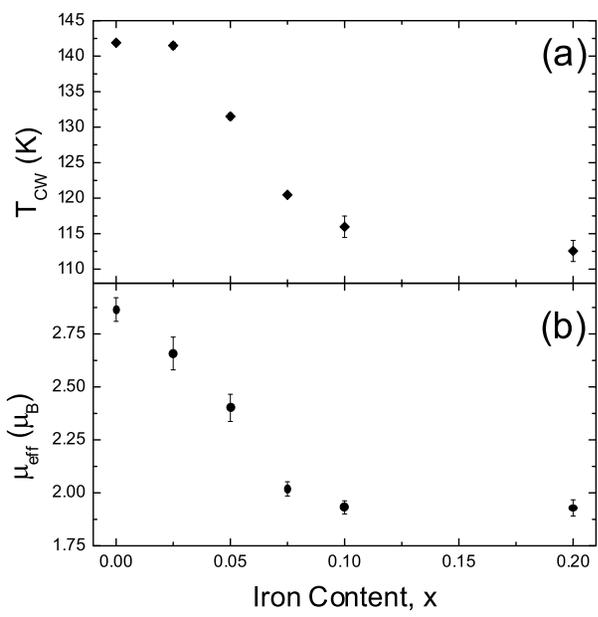

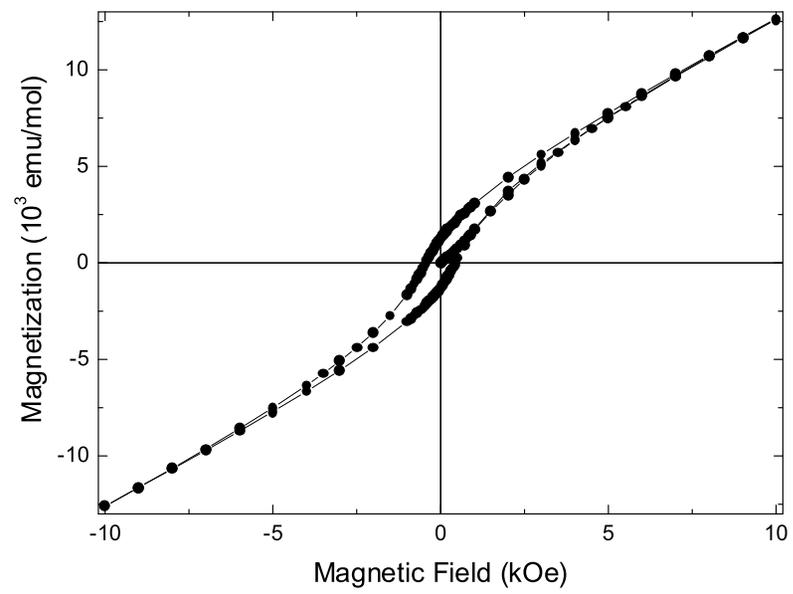

TABLE 1

| x | | 0.0 | 0.025 | 0.05 | 0.075 | .10 | 0.2 |
|---|---|---|---|---|---|---|---|
| | a(Å) | 3.8369(3) | 3.8373(2) | 3.8385(2) | 3.8393(4) | 3.8395(4) | 3.8412(4) |
| | c(Å) | 11.563(3) | 11.557(2) | 11.555(3) | 11.542(3) | 11.539(4) | 11.528(4) |
| | V(Å$^3$) | 170.23(3) | 170.18(2) | 170.25(2) | 170.13(4) | 170.11(4) | 170.11(4) |
| Sr | z | 0.3067(4) | 0.3067(3) | 0.3067(4) | 0.3066(2) | 0.3066(2) | 0.3066(2) |
| Cu | z | 0.1452(2) | 0.1452(3) | 0.1455(2) | 0.1456(3) | 0.1457(3) | 0.1470(3) |
| O(1) | x | 0.0390(1) | 0.0390(2) | 0.0390(2) | 0.0390(2) | 0.0390(1) | 0.0390(2) |
| | z | 0.3335(3) | 0.3335(3) | 0.3337(3) | 0.3338(4) | 0.3339(4) | 0.3347(4) |
| O(2) | z | 0.1295(1) | 0.1295(2) | 0.1297(4) | 0.1295(3) | 0.1295(3) | 0.1295(4) |
| O(3) | x | 0.1140(2) | 0.1140(1) | 0.1140(1) | 0.1140(2) | 0.1140(1) | 0.1140(2) |
| Ru | N(Fe) | - | 0.01(1) | 0.04(2) | 0.06(2) | 0.07(1) | 0.10(1) |
| Cu | N(Fe) | - | 0.02(2) | 0.01(3) | 0.02(2) | 0.03(2) | 0.08(2) |
| | R$_p$ (%) | 5.5 | 6.0 | 5.9 | 5.3 | 5.9 | 5.4 |
| | R$_{wp}$(%) | 7.2 | 7.9 | 7.7 | 7.0 | 6.7 | 7.0 |
| | R$_{exp}$(%) | 5.5 | 6.4 | 6.0 | 5.4 | 6.3 | 5.7 |
| | $\chi^2$(%) | 1.7 | 1.5 | 1.6 | 1.7 | 1.5 | 1.5 |

TABLE 2

| $x$ | 0.0 | 0.025 | 0.05 | 0.075 | 0.1 | 0.2 |
|---|---|---|---|---|---|---|
| Bond Lengths (Å) | | | | | | |
| Ru – O(1) | 1.931(4) | 1.930(5) | 1.927(3) | 1.924(2) | 1.923(3) | 1.912(4) |
| Ru – O(3) | 1.967(4) | 1.968(5) | 1.969(3) | 1.969(5) | 1.969(4) | 1.970(4) |
| Cu – O(1) | 2.182(4) | 2.181(5) | 2.180(3) | 2.177(2) | 2.177(3) | 2.169(4) |
| Cu – O(2) | 1.927(4) | 1.927(5) | 1.928(2) | 1.929(3) | 1.929(3) | 1.931(4) |
| Bond Angles (degrees) | | | | | | |
| Ru-O(3)-Ru | 154.3(2) | 154.3(3) | 154.3(1) | 154.3(2) | 154.3(4) | 154.3(2) |
| Cu-O(2)-Cu | 169.2(2) | 169.2(1) | 169.1(3) | 168.9(2) | 168.8(2) | 168.0(1) |
| ϕ (Cu-O(1)-Ru) | 171.6(2) | 171.6(2) | 171.6(1) | 171.5(2) | 171.5(1) | 171.5(2) |

TABLE 3.

| $x$ | C(emu K/mol) | $T_{CW}$ (K) | $\mu_{eff}$ ($\mu_B$) |
|---|---|---|---|
| *0.0* | 1.03±0.04 | 141.88±0.37 | 2.87±0.06 |
| *0.025* | 0.89±0.05 | 141.49±0.46 | 2.66±0.08 |
| *0.050* | 0.72±0.04 | 131.50±0.42 | 2.40±0.06 |
| *0.075* | 0.51±0.02 | 120.48±0.32 | 2.02±0.03 |
| *0.10* | 0.45±0.02 | 115.96±1.51 | 1.93±0.03 |
| *0.20* | 0.47±0.02 | 112.56±1.48 | 1.93±0.04 |